\documentstyle[11pt]{article}
\def\dinfn{\smallskip Dipartimento di Fisica, Universit\`a di Trento\\
                           and Istituto Nazionale di Fisica Nucleare,\\
                                   Gruppo Collegato di Trento, Italia}
\newcommand{\s}[1]{\section{#1}\renewcommand{\theequation}
        {\mbox{\arabic{section}.\arabic{equation}}}\setcounter{equation}{0}}

\newcommand{\app}[1]{\section{#1}\renewcommand{\theequation}
        {\mbox{\Alph{section}.\arabic{equation}}}\setcounter{equation}{0}}
\renewcommand{\date}[1]{\par\bigskip\par\sl\hfill #1\par\medskip\par\rm}
\newcommand{\pacs}[1]{\smallskip\noindent{\sl PACS number(s):
                       \hspace{0.3cm}#1}\par\bigskip\rm}
\def\babs{\hrule\par\begin{description}\item{Abstract: }\it}
\def\eabs{\par\end{description}\hrule\par\medskip\rm}

\renewcommand{\thanks}[1]{\footnote{#1}}
\newcommand{\preprint}[1]{\hfill{\sl preprint - #1}\par\bigskip\par\rm}
\renewcommand{\title}[1]{\begin{center}\Large\bf #1
\end{center}\rm\par\bigskip}
\renewcommand{\author}[1]{\begin{center}\Large #1\end{center}}
\newcommand{\address}[1]{\begin{center}\large #1 \end{center}}

\begin{document}

\newcommand{\be}{\beta}
\newcommand{\la}{\lambda}
\newcommand{\ep}{\epsilon}
\newcommand{\fr}{\frac}
\newcommand{\reals}{\mbox{${\rm I\!R }$}}
\newcommand{\nats}{\mbox{${\rm I\!N }$}}
\newcommand{\intgs}{\mbox{${\rm Z\!\!Z }$}}
\newcommand{\cam}{{\cal M}}
\newcommand{\caz}{{\cal Z}}
\newcommand{\cao}{{\cal O}}
\newcommand{\cac}{{\cal C}}
\newcommand{\aaa}{\int\limits_{mR}^{\infty}dk\,\,}
\newcommand{\xyz}{\int\limits_{m}^{\infty}dk\,\,}
\newcommand{\bbb}{\left[\left(\frac k R\right)^2-m^2\right]^{-s}}
\newcommand{\ccc}{\frac{\partial}{\partial k}}
\newcommand{\imk}{\xyz [k^2-m^2]^{-s}\ccc}
\newcommand{\fff}{\frac{\partial}{\partial z}}
\newcommand{\iikma}{\aaa \bbb \ccc}
\newcommand{\ddd}{\int\limits_{mR/\nu}^{\infty}dz\,\,}
\newcommand{\eee}{\left[\left(\frac{z\nu} R\right)^2-m^2\right]^{-s}}
\newcommand{\lll}{\frac{(-1)^j}{j!}}
\newcommand{\iinma}{\ddd\eee\fff}
\newcommand{\cah}{{\cal H}}
\newcommand{\nn}{\nonumber}
\renewcommand{\theequation}{\mbox{\arabic{section}.\arabic{equation}}}
\newcommand{\komplex}{\mbox{${\rm I\!\!\!C }$}}
\newcommand{\sip}{\frac{\sin (\pi s)}{\pi}}
\newcommand{\numr}{\left(\frac{\nu}{mR}\right)^2}
\newcommand{\mzs}{m^{-2s}}
\newcommand{\rzs}{R^{2s}}
\newcommand{\abl}{\partial}
\newcommand{\g}{\Gamma\left(}
\newcommand{\zzz}{\int\limits_{\gamma}\frac{dk}{2\pi i}\,\,}
\newcommand{\yyy}{(k^2+m^2)^{-s}\frac{\partial}{\partial k}}
\newcommand{\ikma}{\zzz\yyy}
\newcommand{\ead}{e_{\alpha}(D)}
\newcommand{\sual}{\sum_{\alpha =1}^{D-2}}
\newcommand{\sulnu}{\sum_{n=0}^{\infty}}
\newcommand{\sujnu}{\sum_{j=0}^{\infty}}
\newcommand{\sujeu}{\sum_{j=1}^{\infty}}
\newcommand{\suani}{\sum_{a=0}^i}
\newcommand{\sunueu}{\sum_{\nu =1}^{\infty}}
\newcommand{\suanzi}{\sum_{a=0}^{2i}}
\newcommand{\zend}{\zeta_{1/2}^{lo,\nu}(s)}
\newcommand{\zehnu}{Z_{1/2}^{lo,\nu}(s)}
\newcommand{\zeh}{Z_{1/2}^{lo}(s)}
\newcommand{\aiehnu}{A_i^{1/2,\nu}(s)}
\newcommand{\gdh}{\Gamma\left(\frac 3 2 -k \right)}
\newcommand{\ameehnu}{A_{-1}^{1/2,\nu}(s)}
\newcommand{\aieh}{A_i^{1/2}(s)}
\newcommand{\anehnu}{A_{0}^{1/2,\nu}(s)}
\newcommand{\ameeh}{A_{-1}^{1/2}(s)}
\newcommand{\aneh}{A_{0}^{1/2}(s)}
\renewcommand{\dh}{\left(\frac 3 2 -k \right)}
\newcommand{\amed}{A_{-1}^{\nu ,D}(s)}
\renewcommand{\and}{A_{0}^{\nu ,D}(s)}
\newcommand{\aid}{A_{i}^{\nu ,D}(s)}
\def\beq{\begin{eqnarray}}
\def\eeq{\end{eqnarray}}


\preprint{UTF 354}
\title{Heat-kernel coefficients and functional determinants
\\ for higher-spin fields on the ball}
\author{Klaus Kirsten\thanks{E-mail address:
kirsten@tph100.physik.uni-leipzig.de}}
\address{Universit{\"a}t Leipzig, Institut f{\"u}r Theoretische Physik,\\
Augustusplatz 10, 04109 Leipzig, Germany}
\author{Guido Cognola\thanks{E-mail address:cognola@science.unitn.it}}
\address{\dinfn}


\date{August 1995}
\babs
The zeta function associated with higher-spin fields on the Euclidean
$4$-ball is investigated. The leading coefficients of the
corresponding heat-kernel expansion are given explicitly and the zeta
functional determinant is calculated.
For fermionic fields the determinant is shown to differ for local and
spectral boundary conditions.
\eabs
\s{Introduction}
Motivated by the need to give answers to some fundamental questions in
quantum field theory, during the last years there has been and
continues to be a lot of interest in the problem of calculating the
heat-kernel coefficients and the determinant of a differential
operator, $L$ (see for example
\cite{ramond81,birelldavies82,buchbinderodintsovshapiro92}).
In mathematics the
interest in the heat-kernel coefficients stems, in particular, from
the well-known connections that exist between the heat-equation and
the Atiyah-Singer index theorem \cite{gilkey84}. The knowledge of the
functional determinant allows one to obtain estimates of different
types
\cite{osgoodphillipssarnak88,osgoodphillipssarnak88a,bransonchangyang92}.
All these informations may be obtained by a knowledge of
the zeta function $\zeta_L (s)$ associated with the operator $L$.
For example, for the heat-kernel coefficients $K_n$ associated with
a positive, elliptic, second order differential operator $L$,
one has the equation \cite{seeley67,voros87}
\beq
K_n ={\rm Res}\,\,\left[\zeta_L(s)\Gamma(s)\right]
\left|_{s=\frac{D-n} 2}\right.\label{residuum}
\eeq
with the dimension $D$ of the spacetime. The most appropriate way of
dealing with the determinant of the operator $L$ was introduced by Ray and
Singer \cite{raysinger71} and consists of defining
\beq
\ln\det \,\, L = -\zeta_L '(0).\label{determinant}
\eeq
Especially in recent times, there has been an increasing interest in
these quantities with the $4$-dimensional Euclidean ball as the
underlying manifold and the relevant operator being the Laplacian
respectively the operator describing higher-spin fields
\cite{bek,begk,dowrob,odddow,moscow,ballspin}
\cite{dowker94a,dowker94b,dowker94c,appsdowker95}.

The origin of this interest may be found
in connection with quantum cosmology and
supergravity \cite{deathesposito91,esposito94}
(see also
\cite{kamenshchikmishakov92,barvinskykamenshchikkarmazin92}
\cite{mosspoletti94,louko88,schleich85}).
For example $\zeta (0) $ determines the scaling of the theory, $\zeta
' (0)$ the one-loop effective action.

Very effective schemes have
been developed for the calculation of
these and similar quantities
\cite{bek,begk,dowrob,odddow,moscow,ballspin,appsdowker95}.
Up to now, the mentioned schemes have been applied mainly to the
spin-$0$ case which means to the Laplacian on the Euclidean ball (see
however \cite{moscow}). The aim of this paper is to continue the
consideration of \cite{bek,begk} and to apply the method developed
there to higher-spin fields.
For fermionic fields one has a choice between local and nonlocal
spectral boundary conditions \cite{deathesposito91a}. It was found
that the $\zeta (0)$ values for both boundary conditions agree. Here
we will see, that other properties of the zeta function as for example
$\zeta '(0)$ and thus the effective action are different.

The outline of the paper is as follows. In section 2 we briefly
describe the method developed in \cite{bek} for the calculation of the
heat-kernel coefficients applying it to the spin-$\frac 1 2$ case with
local boundary conditions. The value for $\zeta (0)$ found in
\cite{deathesposito91} is confirmed. We describe how to obtain an
arbitrary large number of the heat-kernel coefficients, summarizing
the first ones in appendix A. The results for spectral boundary
conditions and other spin fields are also summarized in appendix A.
The way to obtain them is identical to the calculation shown for the
spin-$\frac 1 2 $ field with local boundary conditions and so only
some small details are given at the beginning of the appendix A.
In section 3 we direct our interest to the
calculation of the functional determinant. Once more we choose as an
example the spin-$\frac 1 2$ field with local boundary conditions to
apply the formalism developed in \cite{begk}.
The results for the other spins are then relatively easy obtained and
are given without further considerations at the end of section 3.
The results are seen to
differ for local and spectral boundary conditions.
In the conclusions we summarize the main results of our article.

\section{Heat-kernel coefficients for higher-spin-fields
on the 4-ball}
\setcounter{equation}{0}

To exemplify the method, we will choose the spin-$\frac12$ field with
local boundary conditions. Higher-spin fields can be treated in the same
manner, with minor modifications (see the appendix A).
As explained in detail in \cite{deathesposito91}, the eigenvalues of the
Dirac operator with local boundary conditions on the
4-dimensional ball are the solutions of the equation
\beq
J_{n+1}^2(E_{n,j}R)-J_{n+2}^2(E_{n,j}R)=0,
\qquad\qquad n\geq0,
\label{eq:1.1}
\eeq
where each eigenvalue $E_{n,j}$ carries degeneracy $(n+1)(n+2)$. Using
Eq.~(\ref{eq:1.1}) and following the lines of \cite{bek}, the
associated zeta function is written as a contour integral
\beq
\zeta_{1/2}^{lo} (s)=\sulnu (n+1)(n+2) \ikma
\ln[J_{n+1}^2(kR)-J_{n+2}^2(kR)],
\label{eq:1.2} \eeq
where the contour $\gamma$ may be chosen to run counterclockwise
enclosing all solutions of Eq.~(\ref{eq:1.1}) on the real positive axis.
Here we introduced a mass parameter $m$, because it makes the analytical
continuation procedure slightly easier. The limit $m\to 0$ will be taken
at the end. As it stands, the representation (\ref{eq:1.2}) is valid for
$\Re s >2$.
However, in order to determine the heat-kernel coefficients with
higher index we need the properties of $\zeta_{1/2}^{lo}(s)$ in the range
$\Re s < 0$ and
thus we need to perform the analytical continuation to the left domain
of the complex plane.
Before considering in detail the $n$-summation, it is useful to first
proceed with the $k$-integral alone.

The first specific idea is to shift the integration contour and place it
along the imaginary
axis. In order to avoid contributions coming from the origin $k=0$, we
will consider (with $\nu =n+1$) the expression
\beq
\zend = \ikma \ln \left(k^{-2\nu}
[J_{\nu}^2 (kR)-J_{\nu +1}^2(kR)]
\right),\label{eq:1.3} \eeq
where the additional factor $k^{-2\nu}$ in the logarithm
does not change the result, for no additional pole is enclosed.
One then easily obtains
\beq
\zend =\sip \imk \ln \left(k^{-2\nu }
[I_{\nu}^2 (kR)+I_{\nu +1}^2(kR)]
\right)\label{eq:1.4} \eeq
valid in the strip $1/2 <\Re s <1$.

As the next step of our method,
we make use of the uniform expansion
of the Bessel function $I_{\nu} (k)$ and its derivative for $\nu \to
\infty$ as $z=k/\nu$ fixed \cite{abramowitzstegun72}. It turns out, that
things simplify if the Bessel functions in Eq.~(\ref{eq:1.4}) are
rewritten in combinations of Bessel functions with only one index. One
may show using \cite{gradshteynryzhik65}
(see also \cite{deathesposito91}) that
\beq
\zend &=&\sip \iinma \ln\left(z^{-2\nu }
\left[I_{\nu}^{'2} (z\nu)\right.\right.\nn\\
& &\left.\left.\hspace{4cm}
+\left(1+\frac 1 {z^2}\right)I_{\nu}^2(z\nu)
-\frac 2 z I_{\nu} (z\nu ){I'}_{\nu}(z\nu)
\right]\right).\label{eq:1.5} \eeq
Let us now employ the asymptotic expansion.
One has \cite{abramowitzstegun72} \beq
I_{\nu} (\nu z) \sim \frac 1 {\sqrt{2\pi \nu}}\frac{e^{\nu
\eta}}{(1+z^2)^{\frac 1 4}}\Sigma_1
\label{eq:1.6}
\eeq
with $t=1/\sqrt{1+z^2}$ and $\eta =\sqrt{1+z^2}+\ln
[z/(1+\sqrt{1+z^2})]$, furthermore
\beq
I'_{\nu} (\nu z) \sim \frac 1 {\sqrt{2\pi \nu}}\frac{e^{\nu
\eta}(1+z^2)^{\frac 1 4}} z \Sigma_2,
\label{eq:1.7}
\eeq
where we introduced
\beq
\Sigma_1=
1+\sum_{k=1}^{\infty} \frac{u_k (t)}{\nu ^k},\qquad
\Sigma_2=
1+\sum_{k=1}^{\infty} \frac{v_k (t)}{\nu ^k}.\label{eq:1.8}
\eeq
The first few coefficients are listed in \cite{abramowitzstegun72},
higher coefficients are immediate to obtain by using the recursion
relation given also there.

Actually we will need the asymptotics of (see Eq.~(\ref{eq:1.5}))
\beq
\lefteqn
{\ln\left\{ {I'}_{\nu}^2(z\nu ) +\left(1+\frac 1 {z^2}\right) I_{\nu}^2
(z\nu )-\frac 2 z I_{\nu}(z \nu ){I'}_{\nu} (\nu z )\right\}  }\nn\\
& &\sim \ln\left\{\frac{(1+z^2)^{1/2}e^{2\nu\eta}}{2\pi\nu z^2}
\left[\Sigma_1^2+\Sigma_2^2-2t\Sigma_1\Sigma_2\right]\right\}\nn\\
& & =\ln\left\{\frac{(1+z^2)^{1/2}e^{2\nu\eta}}{2\pi\nu
z^2}2(1-t)\right\}\nn\\
& &\qquad+\ln\left\{\frac 1 {2(1-t)}
\left[\Sigma_1^2+\Sigma_2^2-2t\Sigma_1\Sigma_2\right]\right\},
\nn\eeq
and for that reason we introduce the polynomials
\beq
\ln \left\{\frac 1 {2(1-t)}
\left[\Sigma_1^2+\Sigma_2^2-2t\Sigma_1\Sigma_2\right]\right\}
=\sujeu \frac{D_j (t)}{\nu ^j},\label{eq:1.9}
\eeq
which are easily determined by a simple computer program.

Now comes the main idea of our apporach.
By adding and subtracting $N$ leading terms of the asymptotic
expansion, Eq.~(\ref{eq:1.9}), for $\nu \to \infty$, Eq.~(\ref{eq:1.5})
may be split into the following pieces
\beq
\zend = \zehnu +\sum_{i=-1}^N\aiehnu ,\label{eq:1.10}
\eeq
with the definitions
\beq
\zehnu &=& \sip \iinma\left\{\ln\left[z^{-2\nu}\right.\right.\nn\\
& &\left.\times\left( {I'}^2_{\nu} (z\nu )
+\left(1+\frac 1 {z^2}\right) I_{\nu}^2
(z\nu )-\frac 2 z I_{\nu}(z \nu ){I'}_{\nu} (\nu z )\right)\right]\nn\\
& &\left. \qquad-\ln\left[\frac{(1+z^2)^{1/2}e^{2\nu\eta}}{2\pi\nu
z^{2+2\nu}}\right]-\sum_{j=1}^N \frac{D_j
(t)}{\nu^j}\right\}\label{eq:1.11}
\eeq
and
\beq
\ameehnu &=& \sip \iinma \ln \left(z^{-2\nu}
e^{2\nu\eta}\right),\label{eq:1.12}\\
\anehnu &=& \sip \iinma \ln
\frac{(1+z^2) ^{\frac 1 2}(1-t)}{x^2},\label{eq:1.13}\\
\aiehnu &=& \sip \iinma
\left(\frac{D_i (t)}{\nu ^i}\right).\label{eq:1.14}
\eeq
The essential idea is conveyed here by the fact that the representation
(\ref{eq:1.10}) has
the following important
properties. First, by considering the asymptotics of the
integrand in Eq.~(\ref{eq:1.11}) for $z\to mR/\nu$ and $z\to\infty$, it
can be seen that the function
\beq
\zeh =\sunueu \nu (\nu +1) \zehnu\label{eq:1.15}
\eeq
is analytic on the strip $(2-N)/2<\Re s <1$. For this reason, it gives
no
contribution to the residue of $\zeta_{1/2}^{lo} (s)$ in that strip.
Furthermore,
for $s=-k$, $k\in\nats_0$, $k<-1+N/2$, we have $\zeh =0 $ and, thus, it
also yields no contribution to the values of the zeta function at these
points. Together with Eq.~(\ref{residuum}) this
result means that the heat-kernel coefficients are just determined
by the terms $\aieh$ with
\beq
\aieh  =\sunueu \nu (\nu +1)  \aiehnu.\label{eq:1.16}
\eeq
As they stand, the $A_i^{1/2,\nu}(s) $ in
eqs.~(\ref{eq:1.12}), (\ref{eq:1.13}) and
(\ref{eq:1.14}) are well
defined on the strip $1/2<\Re s<1$ (at least). However, the way how to
obtain the analytic
continuation in the parameter $s$ to the whole of the complex plane, in
terms
of known functions, has been explained recently \cite{bek} and will be
exploited further here. Instead of repeating the analysis presented
there, let us only mention that the analogy of the spin-$1/2$ field with
the scalar field obeying Robin boundary conditions might be taken
advantage off. The relevant numbers entering the final result are the
coefficients in the polynomial $D_i (t)$,
\beq
D_i (t) =\sum_{a=0}^{2i}x_{i,a}t^{a+i}.\label{eq:1.17}
\eeq
In terms of these and restricting to the massless case, we found
\beq
\ameeh &=&\frac{\rzs}{2\sqrt{\pi}}\frac{\Gamma\left(s-\frac 1
2\right)}{\Gamma (s+1)}
[\zeta_R (2s -2)+\zeta_R (2s-3)],\nn\\
\aneh &=&-\frac{\rzs}{2\sqrt{\pi}}\frac{\Gamma\left(s+\frac 1
2\right)}{\Gamma (s+1)} [\zeta_R (2s-1)+\zeta_R
(2s-1)],\label{eq:1.18}\\
\aieh &=& -\frac{\rzs}{2\Gamma (s)}[ \zeta_R(-1+i+2s)
+\zeta_R (-2+i+2s)] \nn\\
& &\hspace{3cm} \times \suanzi x_{i,a} \frac{(i+a)\g s+\frac {i+a}
2\right)}{\g 1+\frac {i+a} 2\right)}.\nn
\eeq
At this point, in order to find the heat-kernel coefficients of the
Dirac operator on the ball with local boundary conditions, one has to
use furthermore only
\beq
\zeta_R (1+\epsilon ) &=& \frac 1 {\epsilon} +\cao (\epsilon ^0 ),\nn\\
\Gamma (\epsilon -n )&=& \frac 1 {\epsilon} \frac{(-1)^n}{n!} +\cao
(\epsilon ^0 ).\nn
\eeq
The final results for the relevant residues respectively function values
are listed below for $k\in \nats$. For $\ameeh$ and $\aneh$ we have
\beq
{\rm Res}\,\,A_{-1}^{1/2}\left(\frac 3 2 -k\right)&=&
\frac{(-1)^{k-1}}{(k-1)!}\frac{R^{3-2k}}{2\sqrt{\pi}
\Gamma\left(\frac52-k\right)}\zeta_R \left(1-2k\right),\nn\\
{\rm Res}\,\,A_{0}^{1/2}\left(\frac 3 2 -k\right)&=&
\frac{(-1)^{k-1}}{(k-2)!}\frac{R^{3-2k}}{2\sqrt{\pi}\Gamma\left(\frac 5
2 -k\right)}\zeta_R \left(1-2k\right),
\qquad (\mbox{for }k>2).\nn
\eeq
For the even indices $i$ and for $n=1,...,k-2$ it reads
\beq
{\rm Res}\,\,A_{2n}^{1/2}\left(\frac 3 2 -k\right)
&=&-\frac{R^{3-2k}}{2\Gamma\left(\frac 3 2 -k\right)}
\zeta_R \left(1+2n-2k\right)\nn\\
&&\qquad\times
\sum_{a=0}^{s(k)}x_{2n,2a+1}\frac{(-1)^{k-a-n}(2n+2a+1)}{(k-2-a-n)!
\g\frac32+n+a\right) },\nn
\eeq
where the summation index $s(k)$ for $k\geq 3n+1$ is given by
$s(k)=2n-1$, whereas for $k<3n+1$ one has $s(k)=k-2-n$, and
\beq
{\rm Res}\,\,A_{2k}^{1/2}\left(\frac 3 2-k\right)&=& -\frac{R^{3-2k}}
{4\Gamma\left(\frac 3 2 -k\right)
}\sum_{a=0}^{4k}x_{2k,a}\frac{(2k+a)\g
\frac{3+a}2\right)}{\Gamma\left(\frac a 2 +1+k\right)}.\nn
\eeq
Finally the contributions for odd indices, $n=1,...,k-1$, read
\beq
{\rm Res}\,\,A_{2n-1}^{1/2}\left(\frac 3 2 -k\right)
&=&-\frac{R^{3-2k}}{2\Gamma\left(\frac 3 2 -k\right)}
\zeta_R \left(1+2n-2k\right)\nn\\
&&\qquad\times\sum_{a=0}^{\bar s
(k)}x_{2n-1,2a}\frac{(-1)^{k-1-a-n}(2n+2a-1)}{(k-1-a-n)!
\g\frac12+n+a\right)},\nn
\eeq
with $\bar s (k) =2n-1$ for $k\geq 3n$ and $\bar s (k) =k-n-1$ for
$k<3n$, and
\beq
{\rm Res}\,\,A_{2k-1}^{1/2}\left(\frac 3 2-k\right)&=& -\frac{R^{3-2k}}
{4\Gamma\left(\frac 3 2 -k\right)
}\sum_{a=0}^{4k-2}x_{2k-1,a}\frac{(2k-1+a)\g 1+\frac
a 2\right)}{\Gamma\left(\frac {1+a} 2 +k\right)}.\nn
\eeq
For the function values we found $A_{-1}^{1/2} (0) =-1/120$, $A_0^{1/2}
(0) =1/24$, $A_{-1}^{1/2}(-l) =A_0^{1/2}(-l)=0$ for $l\in \nats$,
furthermore for $n=1,...,k-1$, and $k\in\nats$,
\beq
A_{2n}^{1/2}(1-k) = \frac{R^{2-2k}}{(k-1)!}
\zeta_R \left( 1+2n-2k\right)
   \sum_{a=0}^{s(k)}x_{2n,2a}
\frac{(-1)^{n+a+1}}{(k-n-a-1)!(a+n-1)!},\nn
\eeq
with $s(k) =2n$ for $k\geq 3n+1$, respectively $s(k) =k-n-1$ for
$k<3n+1$, and
\beq
A_{2n-1}^{1/2}(1-k)&=& \frac{R^{2-2k}}{(k-1)!}
\zeta_R \left(-1+2n-2k\right)\nn\\
&&\qquad\qquad\times\sum_{a=0}^{\bar s (k)}x_{2n-1,2a+1}
\frac{(-1)^{n+a+1}}{(k-n-a-1)!(a+n-1)!},\nn
 \eeq
where $\bar s (k) =2n-2$ if $k\geq 3n-1$ and $\bar s (k) =k-n-1$ if
$k<3n-1$. Finally
\beq
A_{2k}^{1/2}(1-k) &=& \frac{(-1)^k
R^{2-2k}} {4(k-1)!} \sum_{a=0}^{4k}x_{2k,a}\frac{(2k+a)\g 1+\frac a
2\right)} {\g 1+k+\frac a 2\right)},\nn
\eeq
and
\beq
A_{2k+1}^{1/2}(1-k) &=& \frac{(-1)^k
R^{2-2k}} {4(k-1)!} \sum_{a=0}^{4k+2}x_{2k+1,a}\frac{(2k+a+1)\g \frac
{3+a} 2\right)} {\g k+\frac {3+a} 2\right)}.\nn
\eeq
A list of the first coefficients is given in the appendix A. Especially
the result given in \cite{deathesposito91},
$\zeta_{1/2}^{lo}(0)=11/360$, is confirmed.

For the spin-$\frac 1 2$ field with spectral boundary conditions as
for the spin-$1$, spin-$\frac 3 2$, and spin-$2$ field exactly the
same method may be employed
(for the conditions analog to (\ref{eq:1.1}) see
\cite{schleich85,deathesposito91a}).
Thus there is no need to present details of the calculation.
Let us only mention, that the $\zeta(0)$ values
agree with the values published already before (for a summary see
\cite{deathesposito91}). Some of the heat-kernel coefficients
corresponding to the values of
the zeta function on the negative axis are summarized in the appendix
A where we also give all necessary elements to perform the
computation. An arbitrary number of coefficients may be obtained
without any problem.

In \cite{ballspin} the sum rule
\beq
\zeta_{3/2}^{sp} (0) -\zeta_{1/2} ^{sp} (0) =2[\zeta_1 (0) -2\zeta_0
(0) ], \label{sumrule}
\eeq
valid also for all coefficients in the heat-kernel expansion, has been
stated. This might be checked for some coefficients using the results
of the present work and of \cite{bek}. Actually, as may be seen in
appendix A, for odd indices Eq.~(\ref{sumrule}) represents two sum
rules, namely one for the terms proportional to $\sqrt{\pi}$ and one
for the terms proportional to $1/\sqrt{\pi}$. The term $1/\sqrt{\pi}$
is absent for the scalar case and so these parts cancel separately for
the other spins.

\s{Functional determinants for higher-spin fields on the 4-ball}

Let us also concentrate in this section on the spin-$\frac12$ field
with local boundary conditions. For higher-spins we only
write down the results at the end of the section.

In order to calculate the functional determinant of the relevant operator
(Laplace, Dirac, ...), Eq.~(\ref{eq:1.10}) is a
very suitable starting point. Using the definition (\ref{determinant})
we have to choose $N=3$ in Eq.~(\ref{eq:1.11})
in order to obtain an analytic representation of
$\zeta_{\frac12}^{lo}(s)$ around $s=0$.
Explicitly the relevant polynomials read
\beq
D_1 (t)  &=&  -\frac 1 4 t +  \frac 1 {12} t^3 ,\nn\\
D_2 (t) &=& \frac 1 8 t^3 +\frac 1 8 t^4 - \frac 1 8 t^5 - \frac 1 8
t^6,\nn\\
D_3(t)  &=& \frac 5 {192} t^3  +
\frac 1 8 t^4  +\frac 9 {320} t^5 - \frac 1 2 t^6  \nn\\
& &\qquad-\frac{23}{64}t^7+\frac38t^8+\frac{179}{576}t^9 .\nn
\eeq
Using these, the contribution of the asymptotic terms to the
functional determinant is found to be
\beq
\frac d {ds} \sum_{i=-1}^3 A_i^{1/2}(s)\left|_{s=0}\right. &=&
-\frac{1597}{15120}-\frac 1 {180} \gamma - \frac 1 {1080} \pi^2  \nn\\
& &-\frac{11}{180}\ln 2 +\frac {11}{180} \ln R-2\zeta_R'(-3) - 3
\zeta_R' (-2) -\zeta _R'(-1).\nn
\eeq
For the part coming from $Z_{1/2}^{lo} (s)$ some additional
calculation is needed.
First one finds
\beq
{Z_{1/2}^{lo,\nu}} '(0) &=&  -\left[\ln \left(  I_{\nu}^2 (z\nu) +
I_{\nu +1}^2 (z\nu)\right) -2\nu\eta - \sum_{n=1}^3 \frac{D_n (t)
}{\nu^n}\right.\nn\\
& &\left.\qquad
-\ln\left(\frac{(1+z^2)^{1/2}}{\pi\nu z^2} (1-t)\right)\right]
\left.\right|_{z=\frac{mR}{\nu} } .\nn
\eeq
In the limit $m\to 0$ this gives
\beq
{Z_{1/2}^{lo,\nu}} '(0) &=&  2\ln \Gamma (\nu +1) +2\nu - 2\nu \ln \nu
-\ln (2\pi\nu )+\sum_{n=1}^3 \frac{D_n(1)}{\nu^n}.\nn
\eeq
Use of the integral representation of $\ln\Gamma (\nu +1)$
\cite{gradshteynryzhik65}, this may be rewritten as
\beq
{Z_{1/2}^{lo,\nu}} '(0) &=&   2\int\limits_0^{\infty}dt\,\,
\left(-\frac t {12} +\frac {t^3}{720} +\frac 1 2 -\frac 1 t +\frac 1
{e^t-1}\right)\frac{e^{-t\nu} } t .
\eeq
The remaining sum may be done using the techniques developed in
connection with the scalar field. As an intermediate result one
obtains
\beq
{Z_{1/2}^{lo}} '(0) &=& 2\lim_{z\to 0}\left[  \frac 1 {360} \Gamma (z+1)
\zeta_R (z+1) + 3 \Gamma (z-2)
\zeta_R (z-2) - 6 \Gamma (z-3) \zeta_R (z-3) \right.\nn\\
& &\qquad\qquad+
\frac 1 {360} \Gamma (z+2) \zeta_R (z+2) -\frac 1 2 \Gamma (z-1)
\zeta_R (z-1) \nn\\
& &\qquad\qquad\qquad\qquad
\left. +\int\limits_0^{\infty}dt\,\, t^z \frac 1 {e^t -1}
\left(\frac d {dt} + \frac{d^2}{dt^2}\right) \frac 1
{t(e^t-1)}\right].\nn
\eeq
This gives
\beq
{Z_{1/2}^{lo}}'(0) = \frac {11}{90} +\frac 1 {180} \gamma +\frac 1 {1080}
\pi^2 +\frac 8 3 \zeta_R ' (-3) +3\zeta_R ' (-2) +\frac 1 3 \zeta_R
(-1).\nn
\eeq
Summing up, we have
\beq
{\zeta_{1/2}^{lo}} '(0) =\frac{251}{15120} -\frac{11}{180} \ln 2
+\frac{11}{180} \ln R +\frac 2 3 \zeta_R ' (-3) -\frac 2 3 \zeta_R '
(-1).\nn
\eeq
This result agrees with that of Apps, reported in \cite{moscow} and
found using a conformal transformation from the $4$-hemisphere.

We retained the dependence on the radius $R$, because the
coefficient of the $\ln R^2$-term is known to be $\zeta (0)$
and this serves as a small check of the calculation

Once more all other cases may be treated in exactly the same way,
so
we list only the final results which are (we use the suffix
${}^{sp}$ to distinguish spectral boundary conditions from the local
ones)
\beq
{\zeta_{1/2}^{sp}}'(0)=-\frac{2489}{30240}+\frac 1 {45} \ln 2
+\frac{11}{180} \ln R +\frac 2
3 (\zeta_R'(-3)-\zeta_R'(-1)),\nn
\eeq
\beq
\zeta_1'(0)=-\frac{6127}{15120}-\frac {29}{45} \ln 2 -\frac{77}{90}
\ln R -\ln \pi
    +\frac 2 3 \zeta_R ' (-3) -\zeta_R ' (-2) -\frac 5 3 \zeta _R '
(-1) ,\nn
\eeq
\beq
{\zeta_{3/2}^{lo}}'(0)=\frac{2771}{15120} -\frac{71}{180}\ln 2 -2\ln\pi
-\frac{289}{180} \ln R +\frac 2 3 \zeta_R'(-3) -\frac{14} 3
\zeta_R'(-1),\nn
\eeq
\beq
{\zeta_{3/2}^{sp}}'(0) = -\frac{27689}{30240} -\frac{59}{45} \ln 2
-\frac{289}{180}\ln R
-2 \ln \pi +\frac 2 3 \zeta_R'(-3) -\frac{14} 3 \zeta_R '(-1), \nn
\eeq
\beq
\zeta_2 '(0) = -\frac{25027}{15120} +\frac{16}{45} \ln 2
-\frac{556}{45}\ln R - 7\ln \pi  +\frac 2 3 \zeta_R '(-3) -\zeta_R '(-2)
-\frac{23} 3 \zeta_R '(-1).\nn
\eeq
We would like to mention, that the contribution coming from
$Z_{1/2}'(0)$ is identical for spectral and local boundary conditions.
The difference is coming only from $A_0 (s)$ to $A_3 (s)$.

This concludes the list of our results for functional determinants of
higher spin fields on the 4-dimensional Euclidean ball.

All results found agree with the recent results of Dowker \cite{ballspin}.

\section{Conclusions}

In this article we applied the approach developed in \cite{bek,begk}
to the case of higher-spin fields on the ball.
We have seen, that the method is very well suited also for these cases
and that no additional complication compared with the scalar field
appears. We found that the $\zeta (0)$ value for the fermionic
fields agree for local and spectral boundary conditions, however,
other properties of the zeta function contained for example in the
heat-kernel expansion and the zeta functional determinant are seen to
differ for the two boundary conditions.
Arbitrary spin-fields and also higher dimensional balls may be treated
along the same lines.

\section*{Acknowledgments}
We wish to thank Stuart Dowker for helpfully providing his results. KK thanks
the Department of Theoretical Physics of the University of Trento for
their kind hospitality at the final stage of this work.

\appendix

\app{Heat-kernel coefficients for higher-spin fields}

In this appendix we present the results for the spin-$1/2$, $1$, $3/2$
and the spin-$2$ fields obeying the indicated boundary conditions.

The $\lambda$ eigenvalues of the operator (Laplace, Dirac,...)
on the 4-dimensional ball for all the fields we are considering
are the solutions of an equation
$F(J_{\nu},J_{\nu+1})=0$, which involves the Bessel functions.
More precisely, depending on the boundary
conditions, one has
\beq
J_\nu(\la)=0
\label{J1}\eeq
or
\beq
J^2_\nu(\la)-J^2_{\nu+1}(\la)=0
\label{J2}\eeq
where $\nu=n+a$ is positive and $n\geq0$
(for convenience here we put $R=1$).
The degeneration of the eigenvalues is a polynomial of second degree
in $\nu$, that is $d_\nu=\alpha_0+\alpha_1\nu+\alpha_2\nu^2$.
The zeta function for the massless case reads
\beq
\zeta(s)=\sum_{n=0}^\infty d_\nu\int_\gamma
\frac{dk}{2\pi i} k^{-2s}
\frac{\partial}{\partial k}
\ln F(J_\nu(k),J_{\nu+1}(k))
\:,\label{}\eeq
and using the method explained in section 2, one finally gets
\beq
\zeta(s)\Gamma(s)&=&\sum_k
\frac{\alpha_0\zeta_R(2s+k-1,a)+
\alpha_1\zeta_R(2s+k-2,a)+
\alpha_2\zeta_R(2s+k-3,a)}{\Gamma(1-s)}\nn\\
&&\qquad\qquad\qquad\qquad\times
\int_{0}^{1} z^{-2s}(t)C_k(t) dt
+\dots
\:,\label{fs}\eeq
where $z(t)$ has been defined in section 2, while
\beq
\frac{d}{dt}\left[\ln z^\beta-\frac1\nu\ln
F(I_\nu(z\nu),I_{\nu+1}(z\nu))\right]
\sim\sum_k\frac{C_k(t)}{\nu^k},
\label{}\eeq
$\be$ being 1 or 2 according to whether $F$ is given by Eq.~(\ref{J1}) or
(\ref{J2}).
The dots in Eq.~(\ref{fs}) stand for the analytic part in the neighbourhood
of $s=\frac{4-n}2$, where the function on the left-hand
side has a pole with residue $K_n$.
In principle, using Eq.~(\ref{fs}), one can compute
$K_n$ up to any order by a simple computer program.
We easily found

\paragraph{Spin 1/2 - Local Boundary Conditions}
\beq
J^2_{\nu}(\la)-J^2_{\nu+1}(\la)=0,
\qquad\qquad\nu=n+1,
\qquad\qquad d_\nu=\nu+\nu^2\nn.
\eeq

\begin{eqnarray}
K_4 &=& \frac{11}{360}\nn\\
K_5 &=& \frac{93\sqrt\pi}{32768}\nn\\
K_6 &=& \frac{1}{308}\nn\\
K_7 &=& \frac{8987\sqrt\pi}{6291456}\nn\\
K_8 &=& \frac{64627}{29099070}\nn
\end{eqnarray}

\paragraph{Spin 1/2 - Spectral Boundary Conditions}

\beq
J_\nu(\la)=0,\qquad\qquad\nu=n+1,
\qquad\qquad d_\nu=2\nu+2\nu^2\nn.
\eeq

\begin{eqnarray}
K_4 &=& \frac{11}{360}\nn\\
K_5 &=& \frac{817}{20160\sqrt\pi} - \frac{35\sqrt\pi}{32768}\nn\\
K_6 &=& \frac{24341}{1153152}\nn\\
K_7 &=& \frac{115069}{2306304\sqrt\pi} - \frac{911\sqrt\pi}{1572864}\nn\\
K_8 &=& \frac{5294503}{133024320}\nn
\end{eqnarray}

\paragraph{Spin 1 (Maxwell) - Dirichlet Boundary Conditions}

\beq
J_\nu(\la)=0,\qquad\qquad\nu=n+2,
\qquad\qquad d_\nu=-2+2\nu^2\nn.
\eeq

\begin{eqnarray}
K_4 &=& -\frac{77}{180}\nn\\
K_5 &=& -\frac{1}{8\sqrt\pi} - \frac{291\sqrt\pi}{32768}\nn\\
K_6 &=& -\frac{50549}{720720}\nn\\
K_7 &=& -\frac{25}{192\sqrt\pi} - \frac{4463\sqrt\pi}{1572864}\nn\\
K_8 &=& -\frac{13099069}{124156032}\nn
\end{eqnarray}

\paragraph{Spin 3/2  - Local Boundary Conditions}
\beq
J^2_{\nu}(\la)-J^2_{\nu+1}(\la)=0,
\qquad\qquad\nu=n+2,
\qquad\qquad d_\nu=-1+\nu+\nu^2\nn.
\eeq

\begin{eqnarray}
K_4 &=& -\frac{289}{360}\nn\\
K_5 &=& \frac{1}{2\sqrt\pi} + \frac{605\sqrt\pi}{32768}\nn\\
K_6 &=& \frac{53}{2772}\nn\\
K_7 &=& -\frac{5}{48\sqrt\pi} + \frac{48155\sqrt\pi}{6291456}\nn\\
K_8 &=& -\frac{12016999}{232792560}\nn
\end{eqnarray}

\paragraph{Spin 3/2 - Spectral Boundary Conditions}

\beq
J_\nu(\la)=0,\qquad\qquad\nu=n+2,
\qquad\qquad d_\nu=-8+4\nu^2\nn.
\eeq

\begin{eqnarray}
K_4 &=& -\frac{289}{360}\nn\\
K_5 &=& -\frac{4223}{20160\sqrt\pi} - \frac{547\sqrt\pi}{32768}\nn\\
K_6 &=& -\frac{32011}{274560}\nn\\
K_7 &=& -\frac{485531}{2306304\sqrt\pi} - \frac{8015\sqrt\pi}{1572864}\nn\\
K_8 &=& -\frac{796491}{4702880}\nn
\end{eqnarray}

\paragraph{Spin 2 (Transverse Traceless Modes)
- Dirichlet Boundary Conditions}

\beq
J_\nu(\la)=0,
\qquad\qquad d_\nu=-8+2\nu^2\nn.
\eeq

\begin{eqnarray}
K_4 &=& -\frac{278}{45}\nn\\
K_5 &=& \frac{7}{4\sqrt\pi} - \frac{1059\sqrt\pi}{32768}\nn\\
K_6 &=& \frac{305699}{360360}\nn\\
K_7 &=& \frac{139}{96\sqrt\pi} - \frac{15119\sqrt\pi}{1572864}\nn\\
K_8 &=& \frac{87568801}{103463360}\nn
\end{eqnarray}

\end{document}